\documentclass{article}

\usepackage{authblk}
\usepackage{graphicx}

\title{Preparing local strain patterns in graphene by atomic force microscope based indentation}

\author[1,*]{P\'{e}ter Nemes-Incze}
\author[2]{Gerg\H{o} Kukucska}
\author[2]{J\'{a}nos Koltai}
\author[2]{Jen\H{o} K\"{u}rti}
\author[3]{Chanyong Hwang}
\author[1]{Levente Tapaszt\'{o}}
\author[4]{L\'{a}szl\'{o} P. Bir\'{o}}

\affil[1]{Centre for Energy Research, Institute of Technical Physics and Materials Science, Nanotechnology Department, 2D NanoFab ERC Research Group, Budapest 1525, POB 49, Hungary}
\affil[2]{Department of Biological Physics, E\"{o}tv\"{o}s University (ELTE), P\'{a}zm\'{a}ny P\'{e}ter s\'{e}t\'{a}ny 1/A, 1117 Budapest, Hungary}
\affil[3]{Korea Research Institute of Standards and Science, Center for Nanometrology, Daejeon 305-340, Republic of Korea}
\affil[4]{Centre for Energy Research, Institute of Technical Physics and Materials Science, Nanotechnology Department, Budapest 1525, POB 49, Hungary}
\affil[*]{corresponding author email: nemes.incze.peter@energia.mta.hu}

\begin{document}

\flushbottom
\maketitle

\openup 0.5 em

\textbf{
	Patterning graphene into various mesoscopic devices such as nanoribbons, quantum dots, etc. by lithographic techniques has enabled the guiding and manipulation of graphene's Dirac-type charge carriers.
	Graphene, with well-defined strain patterns, holds promise of similarly rich physics while avoiding the problems created by the hard to control edge configuration of lithographically prepared devices.
	To engineer the properties of graphene via mechanical deformation, versatile new techniques are needed to pattern strain profiles in a controlled manner.
	Here we present a process by which strain can be created in substrate supported graphene layers.
	Our atomic force microscope-based technique opens up new possibilities in tailoring the properties of graphene using mechanical strain.
}

\thispagestyle{empty}

		Graphene, being a two-dimensional crystal, has an exposed surface which makes it easy to manipulate its atomic and electronic structure \cite{Biro2012,Vozmediano2010}.
		Until recently, focus has been on patterning graphene into nanostructures, mostly with the aim of tailoring its charge transport properties \cite{Stampfer2008,Magda2014,Baringhaus2013}.
		However, in such nanostructures electronic states of the rough edges obscure quantum confinement effects \cite{Bischoff2015}.
		One promising alternative to lithographically cutting graphene, is applying mechanical strain to it.
		Changing the charge transport properties of a material by straining its crystal lattice is not a new idea, it has been realized in the silicon industry with success \cite{Bedell2014}.
		Examining the case of graphene, one finds that in order to achieve significant changes in the band structure, large homogeneous strains in the range of 15-20\%\ need to be applied \cite{Pereira2009a}, making this approach impractical because the maximum failure strain is also of this order \cite{Lee2008a}.
		Furthermore, such a large strain is hard to implement into a working electronic device \cite{PerezGarza2014}.
		However, periodic strain patterns show promise of creating new functionality in graphene devices, through electron wave guiding, pseudo-magnetic fields, valley polarization etc. \cite{Pereira2009,Jones2014,Guinea2009,Hallam2015,Neek-Amal2012b,Yang2014,Carrillo-Bastos2016,Stegmann2015,Milovanovic2016}.
		Since the graphene lattice remains intact in these cases, new physics can be explored without the plague of edge disorder observed in lithographically processed nanostructures.
		Strain engineering of this kind has been attempted by harnessing moir\'{e} patterns \cite{Summerfield2016,Woods2014,Lu2012}, using the different thermal expansion of graphene and its support \cite{Summerfield2016,Woods2014,Lu2012,Tapaszto2012a}, substrate induced rippling \cite{Zang2013,Wang2015} and by placing graphene on a pre-patterned substrate \cite{Hinnefeld2017,Mi2015,Tomori2011,Lee2013b,Reserbat-Plantey2014,Metzger2009,Gill2015}.
		The drawback of these approaches is that the amount of strain induced, as well as its crystallographic orientation is not easily controlled.
		Additionally, there is a need for a transfer step to the pre-patterned substrate which may induce defects in the graphene layer.
		Moir\'{e} patterns that form between the graphene and substrate are limited in their applicability, because the pattern is inherently determined by the alignment and lattice parameters of the two materials.
		Furthermore, with the exception of hexagonal boron nitride \cite{Woods2014,Summerfield2016}, such moir\'{e} patterns are constrained to metallic substrates, making charge transport measurements problematic.
		Until now, no truly versatile method of introducing strain into graphene has been demonstrated.
		Here we show that strain patterns can be prepared in a graphene flake on SiO$_2$, through an AFM indentation approach that combines the ability to \emph{write} tailor made strain profiles, with the possibility to control the crystallographic orientation of the strain.

\section*{Results and Discussion}

\begin{figure}[ht]
\centering
\includegraphics[width = \linewidth]{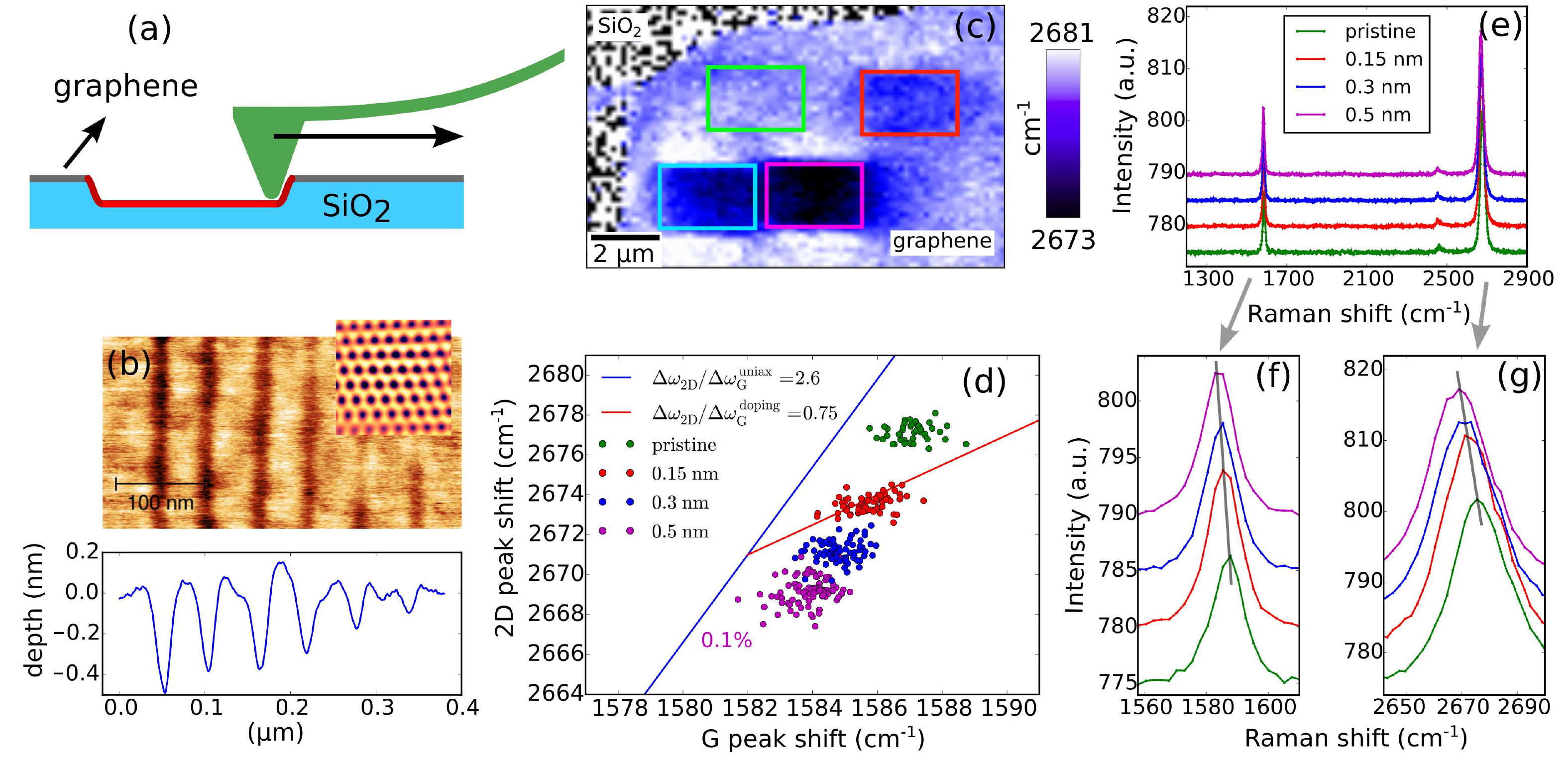}
\caption{
	\textbf{Preparing strain patterns in graphene.}
	\textbf{(a)} AFM indentation patterning scheme.
	\textbf{(b)} AFM topography image and height profile of indentation lines of various depth, prepared by moving the tip along the sample surface (inset: lattice resolved AFM image of graphene flake).
	\textbf{(c)} Raman map of patterned graphene sample. Color scale encodes the position of the 2D peak, obtained by fitting a Lorentz function. The sample contains $2 \times 2.5$ $\mathrm{\mu m^2}$ indent line patterns, having a line spacing of 50 nm, as in b. These line patterns can be easily identified by the increased downshift of the 2D peak wave number and are marked by colored rectangles. The SiO$_2$ substrate areas show up as noise in this image, since the Lorentz fit to the 2D peak fails in this area.
	\textbf{(d)} Correlation plot of the G-2D peak positions measured on line patterns with increasing indentation depth. Colors of the data points correspond to the colors in c. Blue slope corresponds to the ratio of the Gr\"{u}neisen parameters for the 2D and G peaks \cite{Lee2012a}, while the red slope is the shift due to $p$ doping. The maximum average strain relative to the pristine graphene is 0.1\%.
	\textbf{(e)} Raw Raman spectra in a single point measured on the various line patterns.
	\textbf{(f, g)} Plots of the G and 2D peak (colors correspond to the colors used in c). The spectra are offset in intensity with respect to each other for the sake of clarity. Data in this figure was measured, using 532 nm excitation.
}
\label{fig:pattern}
\end{figure}

\begin{figure}[!ht]
\centering
\includegraphics[width = 0.8 \linewidth]{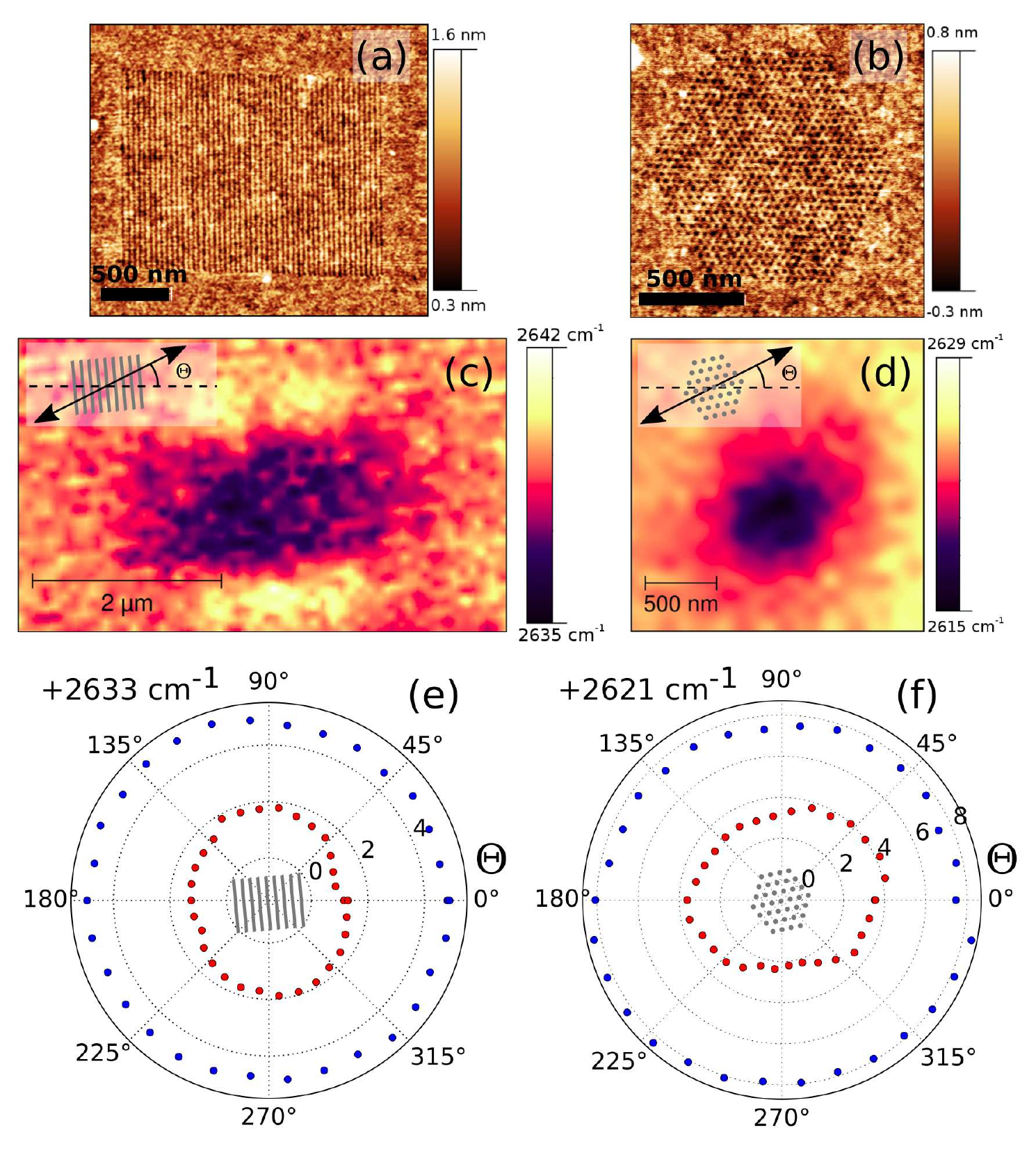}
\caption{
	\textbf{Raman investigation of strain patterns.}
	\textbf{(a, b)} AFM images of an indentation pattern of lines and dots. Dots: nearest neighbor spacing 40 nm. Lines: distance between lines 40 nm.
	\textbf{(c, d)} Raman map of the 2D peak position for each pattern. Inset: definition of the polarizer angle $\theta$ with respect to the pattern. Based on AFM measurements, the orientation of the pattern in the Raman measurement is shown by the sketch of the respective pattern (gray lines and gray dots).
	\textbf{(e, f)} 2D peak shifts of the strain pattern (red) and unperturbed graphene (blue). Gray lines and dots at the center of the plot show the orientation of the patterns with respect to $\theta$. Spectra were measured using 633 nm, linearly polarized excitation. The polarization angle was rotated in $12^\circ$ increments.
}
\label{fig:raman}
\end{figure}

		Our sample system consists of exfoliated graphene flakes on a silicon support, with a 300 nm SiO$_2$ capping layer.
		Being one of the strongest materials known \cite{Lee2008a}, it is natural to assume that in an AFM indentation experiment graphene does not get damaged in the initial phase of the indentation, while the SiO$_2$ substrate can undergo plastic deformation.
		Stopping the indentation before rupture of the graphene occurs, can leave the graphene membrane pinned to the deformed substrate.
		During indentation, the tip is lowered towards the sample surface until a pre-set cantilever deflection (see Methods).
		In the next step, the tip is either retracted or moved along a line on the sample surface (Figure \ref{fig:pattern}a).
		The procedure can be repeated with changing the tip location, resulting in an indentation dot or line pattern (Figure \ref{fig:raman}a,b).
		No significant damage to the graphene has been observed either through AFM or Raman measurements up to a final indentation depth of 1.5 nm.
		With deeper indentation, the rupture of the graphene layer becomes increasingly likely (see Supplementary Figure S1).
		Imaging of the resulting patterns is done using the same tip in tapping mode, unless otherwise noted.
		Importantly, the crystallographic directions of the graphene can be revealed before patterning, by imaging the surface using a softer cantilever (typically 0.1 N/m force constant) in contact mode.
		In this case the frictional forces experienced by the tip are modulated by the atomic lattice (see inset in Figure \ref{fig:pattern}b).

		To determine the magnitude of the strain, we have measured Raman spectroscopy maps of the indentation patterns, with the help of a confocal Raman microscope, using a 532 nm or 633 nm excitation laser.
		If graphene is subjected to tensile strain, both the G and 2D peak positions shift down in wave number, by a factor determined by the respective Gr\"{u}neisen parameter \cite{Mohiuddin2009,Lee2012a}.
		These parameters are in the range of $\partial \omega_{2D} / \partial \epsilon \approx -83 cm^{-1}/ \%$, $\partial \omega_{G^+} / \partial \epsilon \approx -36 cm^{-1}/ \%$ and $\partial \omega_{G^-} / \partial \epsilon \approx -18 cm^{-1}/ \%$ for uniaxally applied strain, as measured by Mohiuddin et al. \cite{Mohiuddin2009}.
		The G peak has two Gr\"{u}neisen parameters, because if the strain has a uniaxial character it will split into two subpeaks called G$^+$ and G$^-$.
		From these parameters it is clear that the 2D peak shows much more shift as a function of strain than the G peak, making possible the detection of strains in the range of 0.01\% \cite{Mohiuddin2009}.
		Because of this property we choose to plot the 2D peak wave number in our Raman maps, to make the strain variations induced by the AFM tip clearer.
		In Figure \ref{fig:pattern}c a plot of the 2D peak position can be seen across a sample area containing $2 \mathrm{\mu m \times 2.5 \mu m}$ arrays of line patterns similar to the one in Figure \ref{fig:pattern}b, each array being composed of 50 indent lines of $2 \mathrm{\mu m}$ length (AFM images of the patterns: supplementary Figure S4).
		As the indentation depth is varied from array to array, from 0.15 nm to 0.5 nm, the downshift in G and 2D peak position becomes stronger, meaning increased strain (see  Figure \ref{fig:pattern}f,g).
		Of course it has to be noted that the strain distribution within the indentation lines will be far from constant \cite{Neek-Amal2012b} and Raman spectroscopy only probes the average of the strain in the graphene inside the laser spot of roughly 500 nm diameter.
		Within these limitations we will quantify the average strain in these structures.
		In Figure \ref{fig:pattern}d we show a correlation plot of the G and 2D peak positions, measured with 532 nm excitation.
		If the G and 2D peak shifts are due to strain effects, their shift is only determined by their respective Gr\"{u}neisen parameters ($\Delta \omega_G$, $\Delta \omega_{2D}$) \cite{Mohiuddin2009}.
		Thus, the measurement points in the correlation plot will lie along a line, the slope of which is determined by the ratio of $\Delta \omega_{2D} / \Delta \omega_{G}$ (blue line).
		This ratio lies within a range of 2.2 to 2.8, depending on the anisotropy in the strain distribution and the crystallographic direction of the strain in the pure uniaxial case \cite{Lee2012a}.
		In addition to strain, the change in the graphene chemical potential can also shift the peak positions.
		If doping effects are significant, the data points will show a deviation from the blue slope.
		If purely doping is the source of the peak shifts the G peak is more strongly affected than the 2D peak and the slope of the line corresponding to it is 0.75, as shown by the red line in Figure \ref{fig:pattern}d.
		Following the evolution of the Raman peak positions with increasing indentation depth, the data points move along the blue line.
		The largest 2D peak shift of 8 cm$^{-1}$, with respect to the unperturbed graphene is observed for the 0.5 nm deep indentation marks, corresponding to an average strain of 0.1\%, using the Gr\"{u}neisen parameter shown above.
		Although splitting of the G peak can be expected, we do not observe this due to the small overall Raman shift.
		Examples of raw Raman spectra, used to create the map and correlation plot in Figure \ref{fig:pattern}c,d can be seen in Figure \ref{fig:pattern}e-g.
		Notice the absence of any disorder induced peak around 1350 cm$^{-1}$, indicating that the number of lattice defects introduced during indentation is negligible.

		Raman spectroscopy also gives us the means to demonstrate that not only can we tune the magnitude of the strain in the patterns, but also to influence the direction and symmetry of the strain.
		In graphene, the crystal momentum of the scattered electron is selected by the polarization of the excitation laser \cite{Gruneis2003}.
		This means that changing the polarization of the laser we can probe the strain in the graphene in various directions.
		Keeping the laser light in the same spot on a line pattern (40 nm line spacing) and a hexagonal dot pattern (40 nm nearest neighbor dot distance), we have measured the dependence of the 2D peak shift with rotating the polarizer of the incident laser beam.
		In Figure \ref{fig:raman}e,f we show a polar plot of the resulting peak shift (red dots), compared with the same measurement performed on the unperturbed graphene next to the patterns (blue dots).
		The data points on the unperturbed graphene form a circle, meaning the average strain distribution within the laser spot is isotropic, as would be expected for graphene on SiO$_2$.
		On the other hand, the measurement on the line pattern shows a 2D peak position that is up to 1 cm$^{-1}$ smaller if the polarization vector of the laser is perpendicular to the indent lines (at $\theta = 15^{\circ}$).
		Thus, the strain has a uniaxial character, being larger in the direction perpendicular to the indent lines \cite{Huang2010}.
		In the case of the dot patterns, the 2D peak shift has a slight hexagonal character, which is aligned with the dot pattern (inset in Figure \ref{fig:raman}d).
		In this case the peak is shifted to higher values by up to 2 cm$^{-1}$, if the polarization is perpendicular to the close packed direction of the indentation dots.
		Therefore, selecting the crystallographic orientation of the pattern, the direction of the strain with respect to the graphene lattice can be set.

\begin{figure}[!h]
\centering
\includegraphics[width = 0.8 \linewidth]{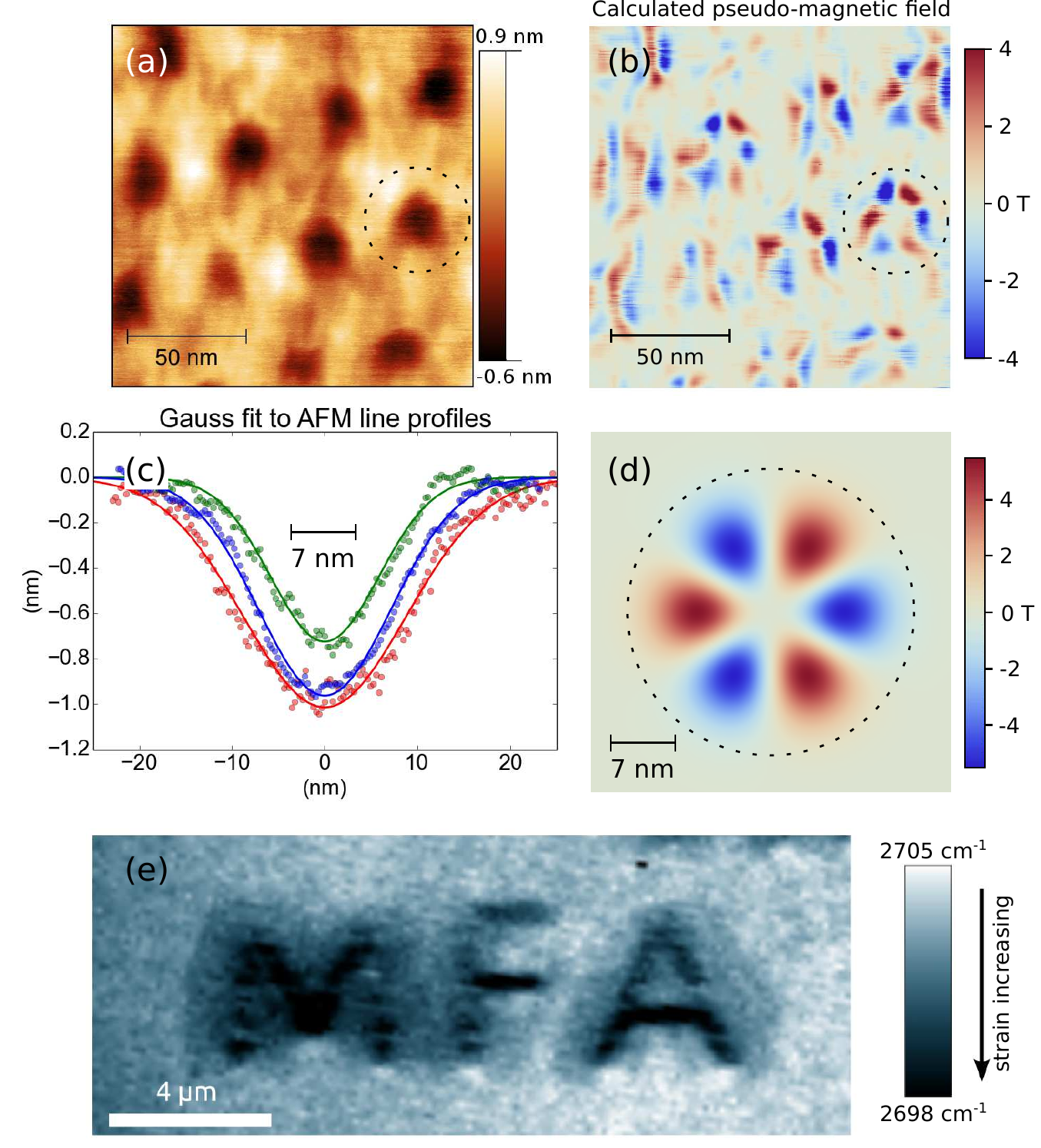}
\caption{
	\textbf{Pseudo-magnetic field and custom strain patterns.}
	\textbf{(a)} AFM image of a dot pattern measured with a sharp AFM tip (2 nm radius of curvature).
	\textbf{(b)} Pseudo-magnetic field pattern calculated from the height profile in (a). Dashed circle marks indentation dot.
	\textbf{(c)} Gaussian fits to AFM height profiles.
	\textbf{(d)} Pseudo-magnetic field calculated for an ideal Gaussian depression with parameters of 1 nm depth and 7 nm variance, similar to the indentation dot marked in (a) and (b).
	\textbf{(e)} Map of the 2D peak position for a strain pattern created in the shape of the initials of our institute. The pattern is composed of a parallel line pattern, such as in Fig. \ref{fig:raman}a with 60 nm line spacing, shaped as the letters MFA. Raman map was measured using 532 nm excitation.
}
\label{fig:mfa}
\end{figure}

		The remarkable observation that graphene stays in the strained configuration after the AFM tip is retracted, leads us to explore the energetics of adhesion.
		The pinning of graphene onto a corrugated substrate can be achieved if the adhesion energy due to van der Waals forces ($E_{vdW}$) is larger than the elastic energy ($E_{el}$) induced in the graphene.
		To be able to compare the two quantities in the present experiment it is necessary to know the exact geometry of the graphene in the pinned configuration.
		AFM probes with a nominal tip radius of curvature of 2 nm have been used to image indentation patterns (see  Figure \ref{fig:mfa}a). 
		Gaussians of the form: $h_{0}(1 - exp(-r^2/2\sigma^2))$, with a variance $\sigma$ in the 7 nm range and depths ($h_0$) from 0.7 to 1 nm, fit the AFM height data very well (Fig. \ref{fig:mfa}c).
		In estimating $E_{el}$ for the present graphene geometry, the bending energy can be safely disregarded, so that the elastic energy is assumed to be dominated by the in-plane stretching of the graphene membrane.
		In this regime we can apply the calculations of Kusminskiy et al. \cite{ViolaKusminskiy2011} for graphene adhered to a Gaussian depression, where the ratio of the Gauss depth to the variance determines the onset of depinning from the substrate.
		For a conservative assumption of graphene-SiO$_2$ adhesion energy \cite{Gao2014,Boddeti2013} of 2 meV/\AA$^2$ the $h_0$/$\sigma < 0.28$ ratio is needed for stable pinning of the graphene to the substrate.
		In the case of the dot patterns prepared here, this ratio is up to 0.14.
		From a mechanical stability point of view, this means that the graphene in the dot patterns is still well within the pinned configuration.
		Estimating the strain from the geometry, one obtains for this dot pattern 0.15\%.
		As the strain in the deformation also scales with $h_0$/$\sigma$, an increase in the possible strain by a factor of 2 could be achieved if AFM tips with smaller tip radius are used for patterning.
		The above calculation assumes that the graphene is adhered by van der Waals forces to the whole surface of the Gaussian shaped hole \cite{ViolaKusminskiy2011}.
		This is a reasonable assumption, since the graphene is pushed into close contact with the support during indentation.
		Therefore, it is expected that the adhesion is improved with respect to exfoliated graphene on SiO$_2$, where the graphene layer is partially suspended \cite{Geringer2009,Georgi2016a}.

		The effect of strain on the orbital motion of electrons in graphene can be described using a vector potential, corresponding to a time reversal symmetric pseudo-magnetic field \cite{Vozmediano2010}.
		This vector potential is of the form: $\vec{A} = \frac{\beta \hbar}{2 a e} (u_{xx}-u_{yy}, -2u_{xy})$, where $\beta \approx 2$, $a$ is the lattice constant, $e$ is the elementary charge and $u_{ij}$ is the strain tensor \cite{Mi2015,Vozmediano2010}.
		The resulting pseudo-magnetic field is given by $B_{\mathrm{ps}} = (\nabla \times \vec{A})_z$, it's effect on graphene's electronic states having been measured previously by scanning tunneling microscopy \cite{Levy2010,Georgi2016a,Lu2012}.
		
		In order to calculate the pseudo-magnetic field induced by the indentation, we need to quantify $u_{ij}$.
		% $u_{i,j} = \left( \partial_i {\mbox{\cal{u}}_j} + \partial_j {\mbox{\cal{u}}_i} + \partial_i h \partial_j h \right) / 2$
		Since, displacements in the $z$ direction (perpendicular to the graphene plane) are much bigger than displacements in-plane, we can safely neglect the in-plane component \cite{Katsnelson2012,Georgi2016a}, resulting in a strain tensor: $u_{ij} = \frac{1}{2} \partial_i h \partial_j h$, where $h$ is the out of plane displacement of the graphene layer.
		We can measure $h$ by AFM topography maps, as long as the AFM tips used for imaging the indentation patterns are much sharper than the ones used to prepare the patterns (see Methods).
		As an example, the AFM topography of an indentation hole pattern (see Fig. \ref{fig:mfa}a) has been used to calculate the strain tensor and the resulting pseudo-magnetic field (Fig. \ref{fig:mfa}b) by numerical differentiation of $h$.
		The resulting pattern of $B_{\mathrm{ps}}$ is largest around the indentation marks (see dashed circle in Fig. \ref{fig:mfa}b) and forms a petal-like structure with alternating positive and negative values of $B_{\mathrm{ps}} \approx$ 4T.
		This flower-like $B_{\mathrm{ps}}$ pattern is characteristic of circularly symmetric deformations \cite{Neek-Amal2012b,Carrillo-Bastos2014,Schneider2015,Georgi2016a} and we can compare this to the $B_{\mathrm{ps}}$ of an ideal Gaussian, because the indentation dots are well fitted by Gaussians (Fig. \ref{fig:mfa}c).
		Fig. \ref{fig:mfa}d shows the calculated $B_{\mathrm{ps}}$ pattern of a Gaussian having a depth of 1 nm and a variance of 7 nm.
		The maximum pseudo-magnetic field in this pattern is 5 T, in good agreement with the $B_{\mathrm{ps}}$ map calculated from the AFM topography data.
		% To include the in plane relaxation of carbon atoms, molecular dynamics calculations are necessary.
		% This has been done by Neek-Amal et al. \cite{Neek-Amal2012b} for centrosymmetric strains and triaxial strain \cite{Neek-Amal2013}.
		% Their results also show a petal structure of the pseudo-magnetic field, similar to continuum approaches \cite{Schneider2015} such as ours.
		% Plugging in the deformation parameters of ref. \cite{Neek-Amal2012b} into our model, we reproduce the pseudo-magnetic field values reported.
		To put the $\sim$4 Tesla pseudo-magnetic field induced by the indentation into perspective, it is instructive to compare it to the pseudo-magnetic field fluctuations resulting from the substrate induced rippling of graphene on SiO$_2$.
		From magneto-transport measurements, such fluctuations were estimated to be in the 1T range \cite{Morozov2006}.
		Therefore, AFM indentation can be used to significantly perturb in a tunable fashion the electronic properties of graphene.

		In summary, scanning probe based techniques have demonstrated remarkable versatility in lithographically cutting nanostructures into graphene \cite{Biro2010,Magda2014}.
		Here we have shown that in an analogous fashion, strain can be induced in SiO$_2$ supported graphene by AFM indentation.
		The crystallographic orientation, magnitude, periodicity of the strain patterns can all be tuned.
		The versatility of the strain patterning technique is demonstrated in  Figure \ref{fig:mfa}e, where we have prepared a strain pattern in graphene showing the initials of our institute.
		These results open up the way to the exploration of tailor made strain profiles in graphene and enable new device concepts, using strain engineering.
		For example, creating periodic strain patterns to realize exotic quantum states, such as a valley ordered ground state \cite{Fu2015}.

\subsection*{Methods}

		\textbf{AFM patterning and imaging details.}
		For the indentation experiments, a Bruker Multimode 8 AFM, equipped with a closed loop scanner, is used.
		For all indentation experiments diamond-like carbon coated silicon AFM probes (Tap300DLC, Budget Sensors) are used, with a nominal force constant of 40 N/m and a tip radius of 15 nm.
		Imaging of the indentation patterns was carried out, using AFM tip having a 2 nm nominal tip radius (SSS-NCH type, NanoWorld)\\
		For the indentation experiments the NanoMan lithography software of Bruker has been used.
		Between indentation steps the tip was moved in tapping mode.
		At the begin of indentation the tip was lowered towards the sample surface with a $z$ velocity of 400 nm/s, until deflection of the cantilever has taken place.
		In the case of the dot patterns the tip was retracted with the same $z$ velocity and moved to a new position for the next indentation step.
		In the case of the line patterns after moving towards the sample the tip was dragged across the surface in contact mode without feedback with a velocity of 200 nm/s.
		Finally the tip was retracted and moved in tapping mode to the new line location.
		The z displacement of the tip was controlled either by setting a cantilever deflection threshold or by moving the tip towards the sample by 40-100 nm.
		The final indent depth was used as a control parameter during indentation experiments because of the variability in cantilever spring constant and tip sharpness.
		The typical cantilever spring constant was 40 N/m, with a tip radius of $\sim$15 nm (Tap300DLC, Budget Sensors).
		However, due to large variability in these parameters, the z movement was incrementally adjusted.
		An indentation experiment was always followed by imaging the patterned location for the onset of plastic deformation of the SiO2.\\
		\textbf{Raman measurements}. Raman measurements were carried out using a Witec 300rsa+ confocal Raman spectrometer, using a 532 nm or 633 nm excitation laser.

% \bibliography{references_main}

% \section*{Acknowledgements}

% 		The work was conducted within the framework of the Korea Hungary Joint Laboratory for Nanosciences through the Korean Research Council of Fundamental Science and Technology P.N.I. and L.P.B. acknowledge the OTKA grant K101599.
% 		G.K. was supported by the UNKP-16-3 National Excellence Program of the Ministry of Human Capacities.
% 		L.T. acknowledges the "Lendület" programme of the Hungarian Academy of Sciences, OTKA grant K108753 and ERC Starting Grant NanoFab2D.
% 		C.H. is supported in part by the Nano-Material Technology Development Program through the National Research Foundation of Korea (NRF) funded by the Ministry of Science, ICT and Future Planning (2012M3A7B4049888).
% 		J. Koltai and J. Kurti acknowledge support from OTKA grants: K-115608, K-108676

\section*{Author contributions statement}

		PN-I designed and carried out the experiments.
		GK has done the calculations.
		CH, LT, JKo, JK\"{u} and LPB reviewed the data.
		JK\"{u}, JKo and LPB supervised the project.
		PN-I wrote the manuscript.
		All authors discussed the data and reviewed the manuscript.

% \section*{Additional information}

% 		\textbf{Supplementary information} accompanies this paper at http://www.nature.com/srep\\
% 		\textbf{Competing financial interests}: The authors declare no competing financial interests.

\end{document}